# History and evidence regarding hydrostatic shock

Michael W. Courtney, U.S. Air Force Academy, 2354 Fairchild Drive, USAF Academy, CO, 80840-6210 Michael.Courtney@usafa.edu

Amy C. Courtney, BTG Research, PO Box 62541, Colorado Springs, CO 80962-2541 amy_courtney@post.harvard.edu

## Abstract

This paper reviews the history and evidence related to remote wounding effects of ballistic pressure waves imparted to tissue by the impact of a bullet. Such remote effects are often referred to as hydraulic or hydrostatic shock. In spite of considerable published evidence and a long history, some medical professionals continue to regard the ability of a bullet to injure tissue that is not directly crushed or stretched as mythical (Jandial R, Reichwage B, Levy M, Duenas V, Sturdivan L. Ballistics for the neurosurgeon. Neurosurgery. 2008:62:472-480.) Early references to these effects date to the 19[th] century; however, compelling experimental support and medical findings in human case studies did not become available until the late 20[th] and early 21[st] century.

Without citing data in support for the claim, the otherwise excellent review paper, *Ballistics for the neurosurgeon*,[1] asserts that "hydrostatic shock" is a "relatively recent myth." However, remote effects of ballistic pressure waves known as "hydrostatic shock" or "hydraulic shock" have considerable support and a long history. Reference to "hydraulic shock" can be found as early as 1898 in an article describing experiments in which fish were killed by a remote pressure mechanism similar to underwater dynamite explosions by firing a rifle into the water within 24 inches or so of the fish.[2] Upon inspection, no easily discernable wound was discoverable on the body, and death was attributed to the remote effects of the pressure wave caused by bullet impacting the water.

In the 1940's, Harvey and co-workers investigated interactions between ballistic waves and tissue in a series of experiments conducted at Princeton University. The fast pressure transients caused by projectiles hitting fluids and fluid filled tissues were detected with piezoelectric pressure transducers and spark shadowgraph photography.[3,4] Without conducting sensitive histological tests or functional tests on living test subjects, this work concluded that the only easily observable injury caused by the observed pressure transients are associated with gas pockets in the body. (The same thinking has long been associated with blast injury.) However, hunters have long attributed instant incapacitation of game animals with remote effects of a rapid pressure transient caused by bullet impact.[5,6]

In 1954, Ochsner reported results of experiments in goats comparing high-speed projectile impacts to the thigh with a small high-explosive charge taped to the same location.[7] Ochsner found it notable that blood transfusions alone could not save these animals and reported an average survival time of 12 hours. He attributed these observations to the pressure changes that had been documented by Harvey.

It was well-documented that tissue could be torn or damaged by temporary cavitation, including neural damage.[8] Documentation of remote damage beyond the reach of the temporary cavity came later, in research reported by groups in Sweden and China. In a series of papers, a Swedish research team reported remote injury to peripheral nerves, spinal cord, and brain using electron microscopy to pigs shot in the thigh.[9,10,11,12] These studies employed high speed pressure sensors implanted in the thigh, abdomen, neck and brain of test animals. The pressure transient was shown to propagate from the impact site to the brain at close to the speed of sound. A Chinese study published in 1990 also reported *in vivo* measurements of fast pressure transients and related remote pressure wave injuries in experiments including pigs and dogs.[13] A later Chinese experiment used sensitive biochemical techniques to detect remote brain injury in dogs shot in the thigh.[14]

Analysis of data relating to of rifle wounds from the Vietnam war described a number of cases of distant wounding including broken bones, abdominal wounding in cases where the bullet did not penetrate the abdominal cavity, a case of lung contusion resulting from a bullet hit to the shoulder, and a case of distant effects on the central nervous system.[15] A case study of a World War II soldier sustaining a handgun wound attributed the much later onset of epilepsy to a hydrodynamic effect.[16] Another case study of a gunshot victim attributed a spinal cord injury to focusing of shock waves remote from the bullet path.[17] A 2007 paper reviewed this and other evidence and predicted that remote brain injury would be common for handgun wounds centered in the chest.[18] Research published in 2009 reported human autopsy results which showed "cufflike pattern haemorrhages around small brain vessels were found in all specimens."[19] This remote brain injury was attributed to the pressure transient caused by the bullet hitting the chest. Easily visible brain hemorrhaging has also been correlated with instant incapacitation in wild animals shot in the chest with much more powerful firearms.[20] Noting similarities between blast and ballistic waves, a recent paper estimated mild traumatic brain injury (TBI) thresholds for the thoracic mechanism of blast induced TBI by analyzing data from ballistic pressure wave and behind armor blunt trauma studies.[21]

A myth is an assertion which has either been disproven by careful experiment or for which there is no historical or scientific evidence in cases where it is reasonably expected. Belief in remote effects of penetrating projectiles may have originated with hunters and soldiers, but their reality is now well established in a broad body of scientific literature, even though the clinical significance for the practicing neurosurgeon might be debatable. Perhaps the clinical significance will become greater with anticipated advancements in detection and treatment of mild TBI.